\documentstyle[aps,prb,twocolumn,epsfig]{revtex}
\begin{document}

\newcommand{\ra}{\rangle}
\newcommand{\re}{\rangle_e}
\newcommand{\rn}{\rangle_n}
\newcommand{\ur}{\uparrow}
\newcommand{\dr}{\downarrow}
\newcommand{\eB}{\mu_BB}
\newcommand{\nB}{g_n\mu_nB}

\twocolumn[\hsize\textwidth\columnwidth\hsize\csname @twocolumnfalse\endcsname

\title
{
 Dynamics of the Measurement of Nuclear Spins in a Solid-State\\
         Quantum Computer
}

\author
{
 Gennady P. Berman$^a$, David K. Campbell$^{a,b}$,
            Gary D. Doolen$^a$, and Kirill E. Nagaev$^d$
}

\address
{
         $^a$Theoretical Division and CNLS, Los Alamos
             National Laboratory, Los Alamos, NM 87545\\
         $^b$Department of Physics, University of Illinois at
             Urbana-Champaign, 1110 West Green St., Urbana, IL
             61801-3080\\
         $^d$Institute of Radio Engineering and Electronics, 
             Russian Academy of Sciences, Mokhovaya St. 11,
             103907 Moscow, Russia
}

\maketitle

\begin{abstract}
{
We study numerically the process of nuclear spin measurement in
a solid-state quantum computer of the type proposed
by Kane by modeling the quantum
dynamics of two coupled nuclear spins on $^{31}$P donors
implanted in silicon.  We estimate the minimum measurement time
necessary for the reliable transfer of quantum information from the
nuclear spin subsystem to the electronic subsystem. We also 
calculate the probability of error for typical values of the external 
noise.
}

\end{abstract}

\pacs{PACS numbers: 03.67.Lx,31.30.Gs}]

\section{ INTRODUCTION}

Recently, a new implementation of a solid-state quantum computer was
proposed by Kane \cite{Kane}. Kane's idea is to realize the quantum qubits 
by using the nuclear
spins of $^{31}$P donors in silicon. These S=1/2 spins are known to
exhibit very long relaxation times because of their weak coupling to
the environment \cite{Waugh}. At low temperatures, when there are no 
thermally
excited electrons in the conduction band,  this contact is mediated
primarily by the magnetic field and the hyperfine interaction with 
{\it s}-electrons localized on the donor. However not all 
{\it s}-electrons are equally important for this mediation. The
electrons in the inner shells are localized within one lattice cell and
are also very weakly coupled to the environment. Of the five
outer (3{\it p}) electrons in a P atom (one more than Si), four form
valence bonds with surrounding Si atoms, while the fifth is given off
to the conduction band to form a loosely bound, hydrogen-like 
{\it s}-state in the field of the positively charged donor. Since the 
motion of this electron is described in terms of the {\it effective}
electron mass, 
which is small in semiconductors, its localization radius can be as large 
as 30 $\AA$. Hence, its state may be controlled by relatively moderate
electric fields. On the other hand, the exchange interaction between
loosely bound electrons located on different donors results in a weak
indirect coupling between their nuclear spins \cite{Slichter}, which
coupling allows the implementation of, e.g., the quantum Control-Not
operation.

\begin{figure}
\centerline{\epsfig{width=3in,file=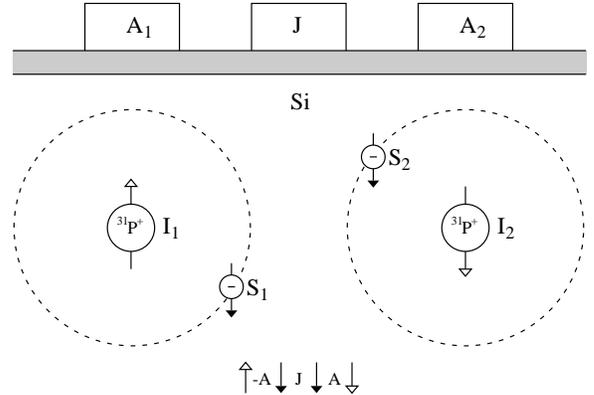}}
\caption{Two $^{31}$P donors in silicon. The nuclear spins are
coupled to the 
outer electrons by the hyperfine interactions, which
can be controlled by the $A$ gates.
The electrons are mutually coupled via an exchange interaction,
which can be controlled by the $J$ gate.}
\label{donors}
\end{figure}

A schematic illustration of the quantum computer proposed by Kane
is shown in
Fig. \ref{donors} for the particular case of two-qubits.
Two $^{31}$P donors are implanted 
in silicon and subjected to an external dc magnetic field
of about $B = 2T$. This creates a
Zeeman splitting of the nuclear spin levels of about $3.5
\times 10^7$ ~Hz and a Zeeman splitting of electron levels of about
$5.6 \times 10^{10}$ ~Hz (much smaller than the splitting between the 
ground and the lowest excited hydrogen-like states of electron, which 
is\cite{enderlein,kohn} 15 meV). The hyperfine coupling
constant for Si:$^{31}$P is 29
MHz. Since this constant is proportional to the probability of finding the
electron near the nucleus, it can be decreased by effects that attract
the electron away from the $^{31}$P nucleus, which in the device
depicted in Fig. \ref{donors} can be accomplished by applying a positive
voltage to the ``gates'' labeled A$_1$ and A$_2$.
Similarly, the exchange
interaction between the electrons located on different donors may be
tuned by applying positive or negative voltage to the J-gate
in Fig. \ref{donors}, thereby
changing the overlap of electron wave functions.

In this article, we do not consider the actual {\it process} of quantum
computation but instead focus on the {\it retrieval} of the result
after the computation has been completed. The weak coupling of
the nuclear spins with their environment, which is essential to
avoid the decoherence that will spoil the quantum computation, makes
this retrieval a highly nontrivial task.
In particular, it cannot be accomplished by means of current
NMR methods, since their sensitivity is as yet
insufficient for detecting the signal from a
single nuclear spin. Recognizing this, Kane proposed a special measurement
procedure based on transferring the information about the {\it spin} state of 
the system to its {\it charge} state, where the result
can be measured. In the following sections, we simulate numerically the 
measurement procedure proposed by Kane and estimate its optimal
duration and minimum probability of error.

\section{ EIGENENERGIES AND EIGENSTATES }

The Hamiltonian of the full system is 
\begin{eqnarray}
H&=&  2\mu_B B      ( \hat{S}_{1z} + \hat{S}_{2z} )
   -2 g_n \mu_n B ( \hat{I}_{1z} + \hat{I}_{2z} ) \nonumber \\
& &  + 4 A_1  \hat{\bf S}_1  \hat{\bf I}_1
   +  4 A_2      \hat{\bf S}_2  \hat{\bf I}_2
   +  4 J        \hat{\bf S}_1  \hat{\bf S}_2,
\label{H}
\end{eqnarray}
where $\hat{\bf S}_i$ and $\hat{\bf I}_i$ are electron and nuclear
spin operators corresponding to donors i=(1,2), $\mu_B$ ($\mu_n$)
is the Bohr (nuclear) magneton, $g_n = 1.13$ is the nuclear
g-factor, $B$ is the external magnetic
field, $A_i$ is the hyperfine interaction constant for nucleus i,
and $J$ determines the strength of the exchange interaction between the
electrons. 

\begin{figure}
\centerline{\epsfig{width=3in,file=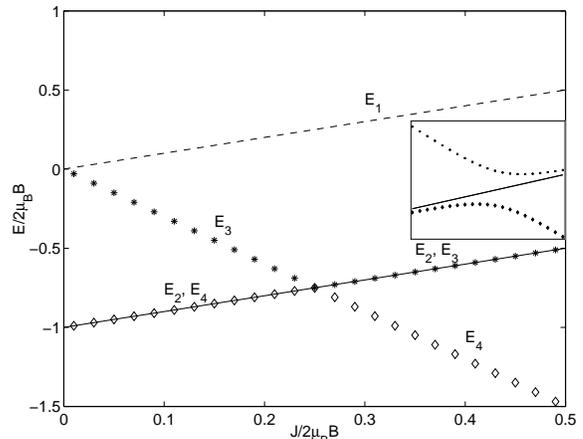}}
\caption{Energy levels of the system of two coupled nuclear and
electron spins versus exchange coupling $J$. The inset
provides an expanded scale view of the 
avoided crossing (``bottleneck'')
formed by  $E_3(J)$ and $E_4(J)$ curves at $J= \eB/2$.}
\label{levels}
\end{figure}

\begin{figure}
\centerline{\epsfig{width=3in,file=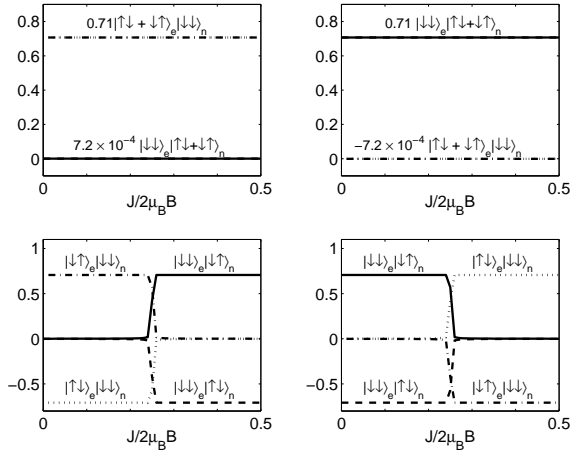}}
\caption{The components of $|E_1\ra$ ,$|E_2\ra$, $|E_3\ra$, and $|E_4\ra$ states 
versus $J$.}
\label{vectors}
\end{figure}

The Hamiltonian $H$ can be represented as a $16 \times 16$ matrix
in the basis of states with definite electron and nuclear spin
projections. Since the total spin projection in the field
direction is conserved, all possible states fall into five
invariant subspaces, corresponding to its values -2, -1, 0, 1, and 2.
In what follows, we
will be interested only in the states with $S_{z\Sigma} + I_{z\Sigma}
= -1$, which are used for measuring of the nuclear spin states.
Hence we may focus on a reduced basis of four states: 
$|\dr\dr\re|\dr\ur\rn$,
$|\dr\dr\re|\ur\dr\rn$,
$|\dr\ur\re|\dr\dr\rn$, and
$|\ur\dr\re|\dr\dr\rn$.
In this reduced basis, the Hamiltonian can be represented
by the $4 \times 4$ matrix
\begin{equation}
{\small  
H
=
\left(
   \begin{array}{cccc}
      J - 2\eB &        0 &     2A_1 &        0 \\
             0 & J - 2\eB &        0 &     2A_2 \\
          2A_1 &        0 & 2\nB - J &       2J \\
             0 &     2A_2 &       2J & 2\nB - J \\
   \end{array}
\right).
\label{4x4}  }
\end{equation}
For simplicity, we henceforth assume
that $A_1 = A_2 = A$. Then the Hamiltonian is symmetric
with respect to donors 1 and 2, and the eigenstates are either
symmetric or antisymmetric with respect to interchanging them. The two 
symmetric states, $|E_1\ra$ and $|E_2\ra$, have total electron and
nuclear spins ${\bf S}_{\Sigma} = 1$ and ${\bf I}_{\Sigma} = 1$.
The corresponding eigenenergies are given by
{\small
\begin{eqnarray}
  &  E_1 = \nB + J - \eB + \sqrt{ (\eB + \nB)^2 + 4A^2 }, &
  \label{E1}\\
  &  E_2 = \nB + J - \eB - \sqrt{ (\eB + \nB)^2 + 4A^2 }, & 
  \label{E2}  
\end{eqnarray}
}
and the corresponding (unnormalized) eigenvectors are
{\small
\begin{eqnarray}
& |E_1\ra =&  -(2\nB + J - E_1)|\dr\dr\re|\dr\ur + \ur\dr\rn
 \nonumber \\
  & & + 2A|\dr\ur + \ur\dr\re|\dr\dr\rn,
  \label{vec1} 
\end{eqnarray}
\begin{eqnarray}
& |E_2\ra=&  -(2\nB + J - E_2)|\dr\dr\re|\dr\ur + \ur\dr\rn
 \nonumber \\
 & & + 2A|\dr\ur + \ur\dr\re|\dr\dr\rn,
 \label{vec2}
\end{eqnarray}
}
The eigenenergies $E_1$ and $E_2$ are linear functions of $J$ (see
Fig. \ref{levels}), and the structure of the corresponding eigenstates
is independent of $J$ (see Fig. \ref{vectors}).

The Hamiltonian $H$ also has two antisymmetric states, $|E_3\ra$ and
$|E_4\ra$, with total spin ${\bf S}_{\Sigma} + {\bf I}_{\Sigma}
= 1$. The eigenenergies $E_3$ and $E_4$ are given by
{\small
\begin{eqnarray}
  &  E_3 = \nB - J - \eB + \sqrt{ (\eB + \nB - 2J)^2 +4A^2 }, &
  \label{E3}\\
  &  E_4 = \nB - J - \eB - \sqrt{ (\eB + \nB - 2J)^2 +4A^2 }, &
  \label{E4}
\end{eqnarray}
}
and the corresponding (unnormalized) eigenvectors are
{\small
\begin{eqnarray}
& |E_3\ra =& -2A|\dr\dr\re|\dr\ur - \ur\dr\rn \nonumber \\
& &  -(2\eB - J + E_3)|\dr\ur - \ur\dr\re|\dr\dr\rn,
 \label{vec3}
\end{eqnarray}
\begin{eqnarray}
& |E_4\ra =& -2A|\dr\dr\re|\dr\ur - \ur\dr\rn \nonumber \\
& & -(2\eB - J + E_4)|\dr\ur - \ur\dr\re|\dr\dr\rn,
 \label{vec4}
\end{eqnarray} 
}

Unlike the energies of symmetric states, $E_3$ and $E_4$ are
non-monotonic functions of $J$ (see Fig. \ref{levels}). If $A$ were
zero, the $E_3(J)$ and $E_4(J)$ curves would intersect at $J =
\eB/2$. However, since these terms have the same symmetry, they
repel to form an avoided crossing (``bottleneck'')
of width $4A$ at the would-be intersection 
point (see Fig. \ref{levels}, inset). The corresponding eigenvectors
also interchange character at this point. That is, as $J$ increases
from $J < \eB/2$ to $J > \eB/2$, $|E_3\ra$ transforms 
from $|\dr\ur - \ur\dr\re|\dr\dr\rn$ with ${\bf S}_{\Sigma} = 0$ and
${\bf I}_{\Sigma} = 1$ into $|\dr\dr\re|\dr\ur - \ur\dr\rn$ with 
${\bf  S}_{\Sigma} = 1$ and ${\bf I}_{\Sigma} = 0$. In other words,
the electron and neutron subsystems exchange their
spins. Simultaneously, $|E_4\ra$ undergoes the inverse transformation from 
$|\dr\dr\re|\dr\ur - \ur\dr\rn$ into $|\dr\ur - \ur\dr\re|\dr\dr\rn$,
with the opposite spin transfer. 

One can thus distinguish between the singlet and triplet states of the
nuclear subsystem as follows. Suppose that after the quantum
computation has been performed with $J = 0$, the electron subsystem is 
in the $|\dr\dr\re$ state. Hence the whole system is either in
$|E_2\ra$ or in $|E_4\ra$ state, depending on whether the nuclear
subsystem is in the $|\dr\ur +\ur\dr\rn$ or $|\dr\ur - \ur\dr\rn$
state. If the exchange parameter is then adiabatically increased to $J 
\gg \eB/2$, the final
electron subsystem is (respectively) a triplet or singlet 
state, thus allowing the information to be transferred from the nuclear 
to the electron spin subsystem. To complete the measurement, one
can distinguish between these electron spin states through the
difference in their charge properties, as described by Kane \cite{Kane}.

\section{ SIMULATIONS OF THE MEASUREMENT DYNAMICS }

To simulate the dynamics of measurement, we solve numerically the equation 
\begin{equation}
\frac{d\rho}{dt}
=
\frac{i}{\hbar}
[\rho, {\cal H}(t) + \delta{\cal H}(t)],
\label{rho}
\end{equation}
where $\rho$ is the $4 \times 4$ density matrix of the
system, ${\cal H}(t)$ is given by (\ref{H}) with $J(t)$ linearly
increasing from 0 to $2\eB$. We allow for a fluctuating correction
\begin{equation}
\delta{\cal H}
=
4 \delta A_1(t)  \hat{\bf S}_1  \hat{\bf I}_1
+ 
4 \delta A_2(t)  \hat{\bf S}_2  \hat{\bf I}_2
\label{dH}
\end{equation}
to account for the voltage noise of the A-gates. The random classical 
quantities $\delta A_1$ and $\delta A_2$ are assumed to have 
correlation time $\tau_c$
much shorter than all the dynamic time scales 
and zero averages. Following Abragam \cite{Abragam}, we
rewrite the equation for the density matrix in the form
\begin{equation}
\frac{d\rho}{dt}
=
-\frac{i}{\hbar}
[{\cal H}, \rho]
-
\frac{\tau_c }{\hbar^2}
\overline
{
  [ \delta{\cal H}(t),
          [
            \delta{\cal H}(t), \rho
          ]
  ]
},
\label{rho2}
\end{equation}
where overlining denotes averaging over realizations of the
fluctuations. We estimate the spectral density of fluctuations of
$A_1$ and $A_2$ using the spectral density of voltage
fluctuations for room-temperature electronics $S_V = 10^{-18}$
~V$^2$/Hz and $d(A/h)/dV \sim 30$ MHz/V (see Ref. \onlinecite{Kane}).
The system is
initially prepared in the $|E_4\ra$ state.

\begin{figure}
\centerline{\epsfig{width=3in,file=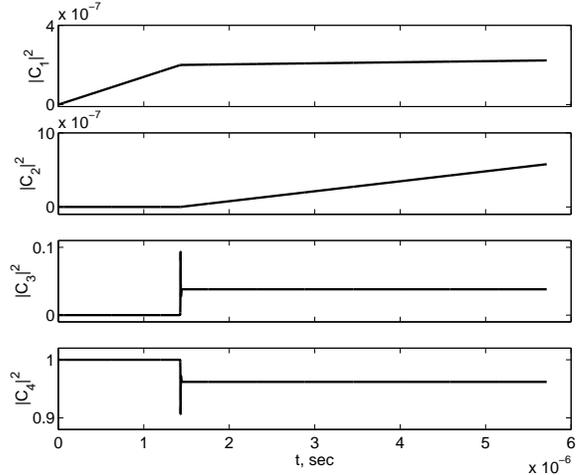}}
\caption{ The time evolution of the occupancies of $|E_1\ra$ - $|E_4\ra$ states 
during the measurement. The duration of measurement is 
$T = 5.7 \times 10^{-6}$
~sec, the hyperfine interaction constant is 
$A/h = 2.9 \times 10^{7}$
~Hz, and the noise spectral density is $S_A/h^2 = 3.5 \times
10^{-3}$ ~Hz.}
\label{dynamics}
\end{figure}

Figure \ref{dynamics} shows the time dependences of occupation
probabilities of the eigenstates $|E_1\ra$ - $|E_4\ra$ during
the measurement process.
In this figure, we have taken
the duration of 
measurement  to be $T = 5.7 \times 10^{-6}$~sec, the hyperfine interaction 
constant to be $A/h = 2.9 \times 10^{7}$ ~Hz, and the noise spectral density 
to be  $S_A/h^2 = 3.5 \times 10^{-3}$ ~Hz. It is readily seen
that increasing $J$ at a finite rate results in a finite probability 
of exciting the system from $|E_4\ra$ into $|E_3\ra$, creating
one source of measurement error. Another
source of error is noise-induced ``escape'' into
the eigenstates $|E_1\ra$ and $|E_2\ra$.
Note that the system cannot be excited into these states in
the absence of noise because of their different symmetry. Since the
fluctuations of $A_1$ and $A_2$ violate the symmetry with respect to
donors 1 and 2, they make this excitation possible. The different
shapes of curves representing $|C_1(t)|^2$ and $|C_2(t)|^2$ are explained by
the ``restructuring'' of the eigenstate $|E_4\ra$ state at $J =
\eB/2$:
the noise-induced
transitions are accompanied by flipping one nuclear and one electron 
spin in opposite directions. The noise also contributes to the
escape to $|E_3\ra$.

\begin{figure}
\centerline{\epsfig{width=3in,file=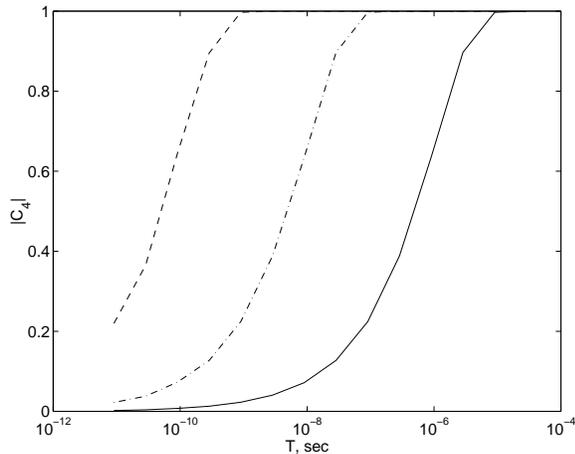}}
\caption{The amplitude of $|E_4\ra$ versus the duration of
measurement for $B = 2$ Tesla and $A/h = 2.9 \times 10^7$ ~Hz
(solid curve). The dash-dotted and dashed curves show the
same dependences for $A/h = 2.9 \times 10^8$ ~Hz
 and $A/h = 2.9 \times 10^9$  Hz, respectively.}
\label{A-curves}
\end{figure}

Figure \ref{A-curves} depicts the dependence of final amplitude of
$|E_4\ra$ on the duration of measurement (solid curve).
For small $T$, 
$|C_4|^2 \to 0$ because the short-time dynamics of the electrons is only
weakly affected by the hyperfine interaction, and the electron subsystem
retains its spin during the course of the measurement. For sufficiently large
$T$, $P_4$ tends to 1 
according to the law $1 - |C_4|^2 \propto \exp(-T/\tau)$. For
representative
values of $B = 2$ Tesla and $A/h = 2.9 \times 10^7$ Hz, 
one obtains $\tau^{-1} = 5.7 \times 10^5$ sec$^{-1}$. 

\begin{figure}
\centerline{\epsfig{width=3in,file=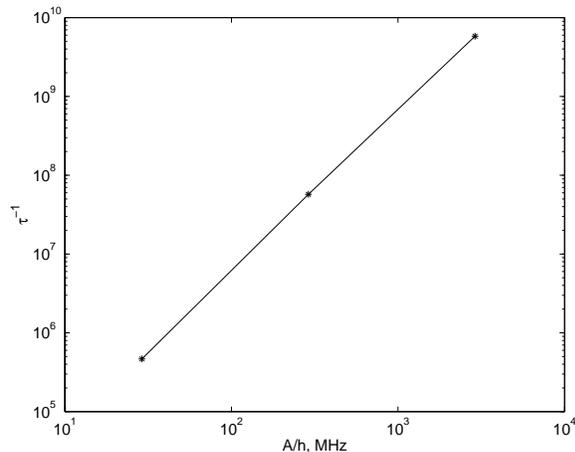}}
\caption{The characteristic time $\tau$ versus the hyperfine interaction constant $A$.}
\label{tau-vs-A}
\end{figure}

To understand why $\tau^{-1}$ is three orders of magnitude
smaller than $A$, we calculated the same curve for $A/h = 2.9 \times
10^8$ Hz and  $A/h = 2.9 \times 10^9$ Hz (dash-dotted and dashed curves).
Figure \ref{tau-vs-A} shows the logarithmic plot of $\tau^{-1}$ vs $A$.

By scaling $A$, we found that $\tau^{-1} \propto A^2 / 2\mu_B B
\hbar$. This result can be readily explained. The width of the
gap between the states
$|E_4\rangle$ and $|E_3\rangle$ is $\Delta E = 4A$ (see
Fig. 2, inset), and
the corresponding interval of $J$ is also of the order of $A$. Hence
the result that $\tau^{-1} \propto A^2$ follows from noting
that $T/\tau \sim \Delta E/\hbar \,\Delta t$, where $\Delta E \propto
A$ and $\Delta t$, which is the time of
passage of the system through the ``bottleneck'' region,
is also proportional to $A$, since $J$ is ramped up linearly
in our simulation.

\begin{figure}
\centerline{\epsfig{width=3in,file=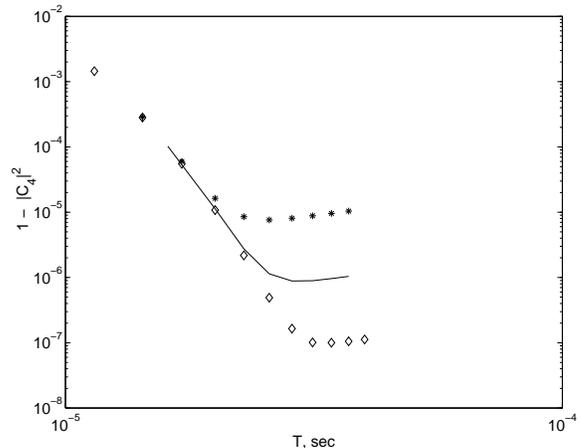}}
\caption{The measurement error, $1 - |C_4|^2$, versus the
  duration of measurement for
different levels of gate-voltage noise: 
$S_A/h^2 = 3.5 \times 10^{-3}$ 
Hz ($\ast$), $3.5 \times 10^{-4}$ Hz (solid line), and $3.5 \times
10^{-5}$ Hz ($\diamond$).}
\label{minima}
\end{figure}

\begin{figure}
\centerline{\epsfig{width=3in,file=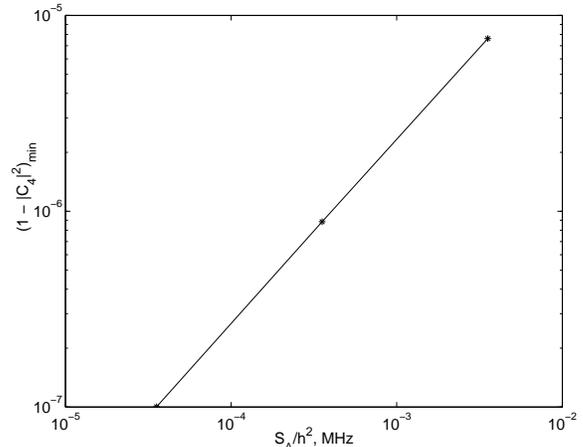}}
\caption{
  The minimum attainable measurement error versus the
  noise spectral density.}
\label{min-vs-S}
\end{figure}

If it were not the gate-voltage noise, the measurement error could be
made vanishingly small just by increasing the duration of
measurement. However, because of noise, the error
passes through a minimum as a function of duration and 
then increases again. The details of behavior of the measurement error 
are shown in Fig. \ref{minima}.
The minimum error $(1 -|C_4|^2)_{min}$ 
is proportional to the noise spectral density $S_A$ (see Fig. 
\ref{min-vs-S}) and is of the order of $10^{-6}$ for typical values of 
noise.    

\section{ ACKNOWLEDGEMENTS }

It is a pleasure to thank Chris Hammel, Bruce Kane, and Denis
Pelekhov for valuable  
discussions. K.E.N. is grateful to the Theoretical Division and the CNLS 
of the Los Alamos National Laboratory for their hospitality. 
This work was supported by the Department of Energy under 
contract W-7405-ENG-36, by the National Security Agency,
and by Linkage Grant 93-1602 from the NATO 
Special Programme Panel on Nanotechnology.


\begin{thebibliography}{99}

\bibitem{Kane} B.E. Kane, Nature {\bf 393}, 133 (1998).
%
\bibitem{Waugh}  J.S. Waugh and C.P. Slichter, Phys. Rev. B {\bf 37},
  4337 (1988).
%
%
\bibitem{Slichter} C.P. Slichter, {\it Principles of Magnetic
    Resonance,} Springer, 1996.
%
\bibitem{enderlein}
R. Enderlein, N.J.M. Horing, {\it Fundamentals of Semiconductor Physics and Devices}, World Scientific, 1997.
%
\bibitem{kohn}
W. Kohn, {\it Solid State Physics}, Vol. 5, (Eds: F. Seitz, D. Turnbull), Academic, New York, 1957.
%
%
\bibitem{Abragam} A. Abragam, {\it The Principles of Nuclear
    Magnetism}, Oxford University Press, 1961.

\end{thebibliography}
\end{document}